\documentclass[fleqn,11pt]{article}
\usepackage{graphicx}
\usepackage{amsfonts}

\newcommand{\bq}{\begin{equation}}            
	\newcommand{\eq}{\end{equation}}            
\newcommand{\bqa}{\begin{eqnarray}}            
	\newcommand{\eqa}{\end{eqnarray}}            
\newcommand{\ba}{\begin{array}} 
	\newcommand{\ea}{\end{array}} 
\newcommand{\bi}{\begin{itemize}} 
	\newcommand{\ei}{\end{itemize}}

\newcommand{\ga}{\gamma}

\newcommand{\nb}{{\nabla}}
\newcommand{\vs}{\vspace{2mm}}

\newcommand{\ra}{\rightarrow}           

\newcommand{\df}{\stackrel{\triangle}{=}}     

\begin{document}
\topmargin=-0.5in

\begin{center}
	{\Large \bf Stability and Convergence of Stochastic Particle Flow Filters}\\[2mm]
\end{center}
\begin{center}
	\begin{tabular}{c  c}
		Liyi Dai & Fred Daum\\
		Raytheon Missiles \& Defense & \hspace{10mm} Raytheon Missiles \& Defense \\
		50 Apple Hill Drive & 235 Presidential Way \\
		Tewksbury, MA 01876 & Woburn, MA 01801 \\
		liyi.dai@raytheon.com & daum@raytheon.com 
	\end{tabular}
\end{center}

\begin{abstract}
In this paper, we examine dynamic properties of particle flows for a recently derived parameterized family of stochastic particle flow filters for nonlinear filtering and Bayesian inference. In particular, we establish that particles maintain desired posterior distribution without the Gaussian assumption on measurement likelihood. Adopting the concept of Lyapunov stability, we further show that particles stay close but do not converge to the maximum likelihood estimate of the posterior distribution. The results demonstrate that stability of particle flows is maintained for this family of stochastic particle flow filters. 
\end{abstract}

{\bf Keywords.} particle flow filters, nonlinear filtering, Bayesian inference, Lyapunov stability

\vspace{0.5cm}

\centerline{May 19, 2021}


\section{Introduction}

Nonlinear filtering plays a critical role in many important applications such as target tracking and Bayesian inference. Particle filtering is a popular method of nonlinear filters but is known to suffer from a problem of “particle degeneracy" \cite{DH2011,DJ,PS,QMG,RAG}. Particle flow filters was proposed in \cite{DH2007} to mitigate the problem of “particle degeneracy" in particle filters. Since then, a number of particle flow filters have been derived in the literature \cite{CL,DH2011,DH2013,DH2015,DHN2010,DHN2016,DHN2018,Khan}. Implementations of particle flow filters for tracking problems in practice have shown that particle flow filters perform remarkably well in practice: Particle flow filters offer orders of magnitude improvements in accuracy and/or speed over conventional extended Kalman filters and particle filters while avoiding “particle degeneracy" and “particle impoverishment" in particle filtering \cite{DHN2010}.

Stability and convergence are important to filtering. Generally speaking, theoretical analysis of stability and convergence for nonlinear filters is challenging. Among results existing in the literature, \cite{RGYU} examined the convergence of extended Kalman filters. A comprehensive overview on the convergence of particle filters is available in the survey paper \cite{CD} and the references therein. Stability and convergence for particle flow filters are yet to be fully addressed. For particle flow filters, an adaptive step-size approach was proposed in \cite{MDD} to mitigate stiffness of underlying stochastic differential equations to ensure numerical stability of particle flows. Numerical stability in solving a particle flow equation was analyzed in \cite{DD2021}. For particle flow filters, there are two aspects of stability and convergence analysis: one is for time-based filtering estimates and the other for particle flows. The latter is the subject of this paper.
 
In this paper, we focus on the stability and convergence of a parameterized family of stochastic particle flow filters that was recently derived in \cite{DD2021}. For this family of particle flows, particles are driven by a diffusion process in which the diffusion matrix acts as a free parameter that stabilizes particle flows. Several particle flows established in the literature are shown to be special cases of this new family. Unbiasedness and consistency of estimates constructed based on the stochastic particle flows were also established under Gaussian assumptions on the prior distribution and measurement likelihood in \cite{DD2021}. The freedom of the diffusion matrix can be exploited to improve filtering performance, for example, to mitigate stiffness of stochastic flow equations \cite{DH2014}. In this paper, we further analyze dynamic properties of the particle flows. We first examine the distribution of particle flows and show that particles indeed maintain the correct posterior distribution. The main focus of this paper is the stability and convergence of particle flows for this new family of particle flow filters, to ensure reliable performance in practice. 

The rest of the paper is organized as the follows. In Section 2, we describe a new parameterized family of stochastic particle flow filters. In \cite{DD2021}, we proved that the new particle flows maintain correct probability distribution under Gaussian assumptions on the prior and the measurement likelihood. In Section 3, we derive the distributions of the flows by relaxing the Gaussian assumption. In Section 4, we examine dynamic properties of the particle flows in the sense of Lyapunov stability.


We use $\mathbb{R}^n$ to denote the real valued $n$ dimensional Euclidean space, $\mathbb{R}=\mathbb{R}^1$, $\mathbb{R}^+$ is the set of non-negative real numbers, and $\mathbb{R}^{n\times m}$ is the real valued $n\times m$ matrix space. An identity matrix is denoted by $I$. The superscript $T$ denotes the transpose of a vector or matrix. The trace of a matrix $A$ is $tr(A)$. 
For a random variable $x\in \mathbb{R}^n$, its mean is $E[x]$. For a scalar function $f(x): \mathbb{R}^n\ra \mathbb{R}$, its gradient is $\nb_x f(x)=[\partial f/\partial x_1, \partial f/\partial x_2, ..., \partial f/\partial x_n]^T\in\mathbb{R}^n$,
and its divergence is $div(f) = \sum_{i=1}^n\partial f/\partial x_i\in\mathbb{R}$.

\section{Stochastic Particle Flow Through Homotopy}
Assume a given probability space, on which $x\in \mathbb{R}^n$ is a $n$-dimensional random variable and $z\in \mathbb{R}^d$ is a $d$-dimensional measurement of $x$. Let $p_x(x)$ denote the prior probability density function of $x$ and $p_z(z|x)$ the likelihood of a measurement $z$ conditioned on $x$. The Bayes' Theorem states that the posterior conditional density function of $x$ for a given measurement $z$, $p_x(x|z)$, is given by\footnote{Particle flow filters are constructed for sequential state estimation, updating state estimate as new data is collected. To keep notations as simple as possible, in this paper we focus on one-step Bayesian estimation which can be applied to filtering or inference problems. For multi-step sequential filtering, the Bayes' Theorem is as the following \cite{DH2007,Jaz}
\[
p(x,t_k|Z_k)=p(z_k|x,t_k)p(x,t_k|Z_{k-1})/p(z_k|Z_{k-1})
\]
in which $z_k$ is the $k$-th measurement at time $t_k$, $Z_k=\{z_1,z_2,...,z_k\}$, and $p(z_k|x,t_k)$ is the probability density of measurement $z_k$ at time $t_k$ conditioned on $x$. The probability density functions $g(x)$, $h(x)$, and $p(x)$ in (\ref{pgh}) need to be replaced, respectively, with the following
\[ 
p(x)=p(x,t_k|Z_k), g(x)=p(x,t_k|Z_{k-1}), h(x)=p(z_k|x,t_k).
\]
The rest of discussion follows.
}
\bq
p_x(x|z) = \frac{p_x(x)p_z(z|x)}{p_z(z)}
\label{bayes}
\eq
in which $p_z(z)=\int_x p_z(z|x)p_x(x)dx$ is the normalization factor. Without loss of generality, it is assumed throughout this paper that all probability density functions exist, second order continuously differentiable, and are non-vanishing everywhere. Those assumptions are stronger than we
need but are helpful to maintaining clarity of discussion without resorting to complex notations.
For simplicity, we denote
\bq p(x)=p_x(x|z), g(x) = p_x(x), h(x) = p_z(z|x).
\label{pgh}
\eq

In the particle flow framework, we define a new conditional probability density function as the following 
\bq
p(x,\lambda) = \frac{g(x)h^\lambda(x)}{\Gamma(\lambda)}
\label{homotopy}
\eq
for all $\lambda\in[0, 1]$. In (\ref{homotopy}), $\Gamma(\lambda)$ is the normalization factor so that $p(x,\lambda)$ remains a probability density function for all $\lambda\in[0, 1]$. It's clear from (\ref{homotopy}) that 
\[    p(x,0) = g(x), p(x,1)=p(x). \]
In other words, $p(x,0)$ is the density function of the prior distribution and $p(x,1)$ is that of the posterior distribution. Therefore, the mapping $p(x,\lambda): \mathbb{R}^+\times [0, 1]\longrightarrow \mathbb{R}^+$ in (\ref{homotopy}) defines a homotopy from $g(x)$ to $p(x)$. By taking the natural logarithm on both sides of (\ref{homotopy}), we obtain
\bq
\log p(x,\lambda) = \log g(x)+ \lambda \log h(x) -\log \Gamma(\lambda).
\label{loghom}
\eq
Recall that a major problem with particle filters is “particle degeneracy" \cite{DH2007,DJ,PS,RAG}. To mitigate this problem, particle flow methods move (change) particles (i.e., samples of $x$) as a function of $\lambda$, $x(\lambda)$, so that (\ref{homotopy}), or equivalently (\ref{loghom}), is always satisfied as $\lambda$ changes from $0$ to $1$. The probability density function of $x(\lambda)$ is $p(x,\lambda)$ for all $\lambda\in [0,1]$. The value of $x(\lambda)$ at $\lambda=1$ is used for estimation in problems such as filtering or Bayesian inference. It turns out that there exists much freedom in the choice of $\{x(\lambda), \lambda\in[0, 1]\}$ \cite{DH2015}.
The $x(\lambda)$ could be driven by a deterministic process as in the Exact Flow \cite{DH2015,DHN2010}, or by a stochastic process as in stochastic flows \cite{DH2013,DHN2016,DHN2018}. In this paper, we focus on stochastic flows in which $x(\lambda)$ is driven by the following stochastic process
\bq
dx = f(x,\lambda)d\lambda+q(x,\lambda)dw_{\lambda}
\label{flow}
\eq
where $f(x,\lambda)\in \mathbb{R}^n$ is a drift function, $q(x,\lambda)\in \mathbb{R}^{n\times m}$ is a diffusion matrix, and $w_{\lambda}\in \mathbb{R}^m$ is a $m$ dimensional Brownian motion process in $\lambda$ with $E[dw_\lambda dw_\lambda^T ]=\sigma(\lambda)d\lambda$. The stochastic differential equation (\ref{flow}) is a standard diffusion process \cite{Jaz}. Note that $\{x(\lambda), \lambda \in [0, 1]\}$ is a stochastic process in $\lambda$, not in time. For clarity, we drop its dependence on $\lambda$ but add the dependence back when it is beneficial to emphasize its dependence on $\lambda$. Without loss of generality, we assume that $\sigma(\lambda)=I_{m\times m}$, and denote
\[ Q(x,\lambda) = q(x,\lambda)q(x,\lambda)^T \in \mathbb{R}^{n\times n}. \]
The matrix $Q(x,\lambda)=[Q_{i,j}]$ is always symmetric positive semi-definite for any $x$ and $\lambda$.

Our goal is to select $f(x,\lambda)$ and $q(x,\lambda)$ (or equivalently $Q(x,\lambda)$) such that (\ref{loghom}) is maintained  for the particle $x(\lambda)$ driven by the stochastic process (\ref{flow}) for all $\lambda\in [0, 1]$. To that end, we start with the following necessary condition.
\vs

{\sc Lemma 2.1.} \cite{CL,DHN2018} Assume that all derivatives exist and are continuous. For the particle flow $x(\lambda)$ defined in (\ref{flow}), a necessary condition for its density function to be $p(x,\lambda)$ for all $\lambda\in [0, 1]$ is that $f(x,\lambda)$ and $Q(x,\lambda)$ satisfy the following condition
\bq
\nb_x\frac{\partial \log p}{\partial\lambda} = - \nb_x div(f)-(\nb_x\nb_x^T\log p)f 
-(\nb_x^T f)(\nb_x\log p)+\nb_x[\frac{1}{2p}\sum_{i,j}\frac{\partial^2 (pQ_{i,j})}{\partial x_i\partial x_j}] 
\label{cond1}
\eq 
for all $x$ and $\lambda\in[0, 1]$. For simplicity and without causing confusion, in (\ref{cond1}) and for the rest of discussion in this paper, we omit all variables involved.
\vs

A new parameterized family of stochastic flows was recently derived in \cite{DD2021}. This new family of particle flows is rather general and takes several particle flows existing in the literature as special cases. The form of this family of flows is described in the following Theorem 2.1 in which $Q$ serves as a free parameter.
\vs

{\sc Theorem 2.1.} \cite{DD2021} {\em Assume that
\begin{description}
\item[(A1)] $\nb_x \log g$ and $\nb_x \log h$ are linear in $x$,
\item[(A2)] $\nb_x \nb_x^T \log p$ is non-singular for all $\lambda \in [0, 1]$.
\end{description}
Then for any positive semi-definite matrix $Q(\lambda)$, independent of $x$, (\ref{cond1}) is satisfied by the following choice of $f$.
\bq
f = (\nb_x \nb_x^T \log p)^{-1}[-\nb_x\log h+K(\nb_x \nb_x^T \log p)^{-1}\nb_x\log p],
\label{f1}
\eq
\bq
K(\lambda) = \frac{1}{2}(\nb_x \nb_x^T \log p)Q(\lambda)(\nb_x \nb_x^T \log p)+\frac{1}{2}(\nb_x\nb_x^T\log h).
\label{q1}
\eq
}

Note that $\nb_x\nb_x^T\log p$ and $\nb_x\nb_x^T\log h$ are Hessian matrices of second-order continuously differentiable probability density functions, and thus are symmetric. The diffusion matrix $Q$ is symmetric by definition. Consequently, the matrix $K\in \mathbb{R}^{n\times n}$ is always symmetric by construction. The freedom of $Q$ can be utilized to improve the dynamic properties of the flow (\ref{flow}), thus the performance of nonlinear filtering. For example, $Q$ may be chosen to mitigate a potential issue of stiffness in the flow (\ref{flow}) \cite{DH2014}.

\section{Distribution of Stochastic Particle Flows}
The condition (\ref{cond1}) is a necessary condition only. It is proved in \cite{DD2021} that the distribution of $x(\lambda)$ is indeed $p(x,\lambda)$ under Gaussian assumptions on the prior $g$ and the measurement likelihood $h$. In this section, we establish that the flow $x$ defined by (\ref{flow}), (\ref{f1})-(\ref{q1}) has desired probability density function $p(x,\lambda)$ without the Gaussian assumption on $h$.

Under the assumption (A1), $\nb_x\log g$ and $\nb_x\log h$ are linear in $x$. We may write 
\bq
\log g = \frac{1}{2}x^TA_gx+b_g^Tx+c_g,
\label{logg}
\eq
\bq
\log h = \frac{1}{2} x^TA_hx+b_h^Tx+c_h
\label{logh}
\eq
where $A_g, A_h\in \mathbb{R}^{n\times n}, b_g, b_h\in \mathbb{R}^n, c_g, c_h\in \mathbb{R}$ are constant matrices, vectors, and scalars, respectively, that are not functions of either $x$ or $\lambda$.
Consequently, $\log p$ has the form
\[
\log p = \log g+\lambda\log h-\log \Gamma(\lambda)
 = \frac{1}{2}x^T(A_g+\lambda A_h)x+(b_g+\lambda b_h)^Tx+c_g+\lambda c_h-\log \Gamma(\lambda).
\]
The gradient and Hessian of $\log h$ and $\log p$ are, respectively
\[
\nb_x\log h = A_hx+b_h, 
\nb_x\nb_x^T\log h = A_h,
\]
\[
\nb_x\log p = (A_g+\lambda A_h)x+b_g+\lambda b_h,
\nb_x\nb_x^T\log p = A_g+\lambda A_h.
\]
The matrices $A_g$ and $A_h$ are Hessian matrices of second-order continuously differentiable probability density functions, and thus are symmetric. The assumption (A2) says that $A_g+\lambda A_h$ is non-singular for all $\lambda\in [0, 1]$. In this case, $p(x,\lambda)$ in (\ref{loghom}) is Gaussian distributed with mean $x_\mu(\lambda)$ determined by setting $\nb_x\log p|_{x=x_{\mu}} = 0$, which gives
\bq
x_\mu(\lambda)=-(A_g+\lambda A_h)^{-1}(b_g+\lambda b_h),
\label{xmu}
\eq
and its covariance matrix $P_\mu(\lambda)$ is given by
\bq
P_\mu(\lambda) = -(\nb_x\nb_x^T\log p)^{-1} = - (A_g+\lambda A_h)^{-1}.
\label{Pmu}
\eq
On the other hand, under the assumption (A1), the flow $x$ is governed by the following linear stochastic process
\bq
dx=(Ax+b)d\lambda + qdw_{\lambda}
\label{linear}
\eq
in which
\[ 
A=\nb_x f = (\nb_x\nb_x^T\log p)^{-1}(-\nb_x\nb_x^T\log h+K),
 b=f-Ax.
\]
For the linear stochastic differential equation (\ref{linear}), the mean $\bar{x}$ and covariance matrix $P$ of $x$ satisfy, respectively, \cite{Arn}
\bq
\frac{d \bar{x}}{d\lambda}=A\bar{x}+b,
\label{mean}
\eq
\bq
\frac{d P}{d\lambda} = AP+PA^T+Q.
\label{cov}
\eq
Furthermore, we know from the theory of linear stochastic differential equations that the probability density function of $x$  (\ref{linear}) is Gaussian for each $\lambda$ \cite{Arn}. To prove that the flow defined in (\ref{flow}), (\ref{f1})-(\ref{q1}), or equivalently (\ref{linear}), has desired posterior density function $p(x,\lambda)$ in (\ref{loghom}), we only need to verify that $x_\mu$ and $P_\mu$ defined in (\ref{xmu})-(\ref{Pmu}) indeed satisfy (\ref{mean}) and (\ref{cov}), respectively.
\vs

We first consider the mean $x_\mu$. For notational convenience, denote $S(\lambda)=\nb_x\nb_x^T\log p$.
For $\bar{x}=x_\mu$, using the equality for differentiating an invertible matrix $B(\theta)$, $dB^{-1}/d\theta=-B^{-1}(dB/d\theta)B^{-1}$,  the left hand side (LHS) of (\ref{mean}) is
\[
\textrm{LHS of (\ref{mean})} =\frac{d}{d\lambda}[-S^{-1}(b_g+\lambda b_h)]
=-\frac{dS^{-1}}{d\lambda}(b_g+\lambda b_h)-S^{-1} b_h
\]
\bq
=S^{-1}A_hS^{-1}(b_g+\lambda b_h)-S^{-1}b_h.
\label{mean_LHS}
\eq
Note that $x_\mu$ satisfies $\nb_x\log p |_{x=x_\mu} = 0$ by construction. Setting $\bar{x}=x_\mu$, the right hand side (RHS) of (\ref{mean}) becomes
\[
\textrm{RHS of (\ref{mean})} = f |_{x=x_\mu}
=S^{-1}(-\nb_x\log h+KS^{-1}\nb_x\log p)|_{x=x_\mu}
=S^{-1}(-\nb_x\log h)|_{x=x_{\mu}}
\]
\[
=S^{-1}[-(A_hx_{\mu}+b_h)]
=-S^{-1} A_h [-S^{-1}(b_g+\lambda b_h)]-S^{-1}b_h
=S^{-1} A_h S^{-1}(b_g+\lambda b_h)-S^{-1}b_h 
\]
\[
= \textrm{LHS of (\ref{mean})}
\]
according to (\ref{mean_LHS}).

For the covariance matrix $P_\mu$, we have $P_\mu=-S^{-1}$. Therefore,
\bq
\textrm{LHS of (\ref{cov})} = \frac{d P_{\mu}}{d\lambda}=S^{-1}A_hS^{-1}
\label{cov_LHS}
\eq
and
\[
\textrm{RHS of (\ref{cov})} =AP_\mu+P_\mu A^T+Q
 =S^{-1}(-A_h+K)P_\mu+P_\mu (-A_h+K^T)S^{-1}+Q
\]
\bq
=S^{-1}(2A_h-K-K^T)S^{-1}+Q.
\label{rhs_1s}
\eq
Substituting the form of $K$ (\ref{q1}) into (\ref{rhs_1s}), also noticing the equality (\ref{cov_LHS}), we obtain
\[
\textrm{LHS of (\ref{cov})} 
=S^{-1}A_hS^{-1}=\textrm{RHS of (\ref{cov})}.
\]
To summarize, we have verified that the flow $x$ defined by (\ref{flow}), (\ref{f1})-(\ref{q1}) indeed has the posterior probability density function $p(x,\lambda)$ (\ref{flow}) for all $\lambda \in [0, 1]$ for any symmetric positive semi-definite matrix $Q$ as long as $Q$ is not a function of $x$. 

This derivation is simpler than that in \cite{DD2021} and does not require Gaussian assumption for the measurement likelihood $h$. For example, $A_h$ may be singular or we may even have $A_h=0$, e.g., exponential distribution. Nevertheless, the assumption (A2) implicitly requires that $A_g$ is non-singular, i.e., the distribution of the prior $g$ is Gaussian. Gaussian distribution is often used as an effective approximation in practice.

\section{Dynamical Stability of Particle Flows}

In this section, we examine dynamic properties of the flow $x(\lambda)$ (\ref{flow}) as a function of $\lambda$ over $[0, 1]$, with $f$ defined in (\ref{f1}). In particular, we are interested in the dynamics of $\log p$ and $\nb_x\log p$. Questions we are interested in answering include whether the flow is stable and convergent, and in what sense. Toward that end, we need to address a couple of issues. First, the drift function $f$ in (\ref{flow}) is $\lambda$-varying. The usual eigenvalue based approach to stability analysis does not directly apply to this type of stochastic differential equations. As we shall show later in this section, the concept of Lyapunov stability in nonlinear system analysis is particularly suitable \cite{Kha, Mao,Sha}. Second, traditional concepts of stability for dynamical systems  characterize dynamic behaviors as time goes to infinity. For particle flows, $\lambda$ is limited to $[0, 1]$. In \cite{DD2021}, we adopted concepts of finite time stability to characterize the stability of numerical evaluation of the flow (\ref{flow}), which is important for the implementation of particle flow filters because a numerical solution of (\ref{flow}) is needed to construct state estimate and estimation confidence. However, concepts of finite time stability are mostly concerned with the final state of a flow $x$ at $\lambda=1$ \cite{AACC2006}. The definitions themselves do not reveal much about how $x$ reaches its final state at $\lambda=1$. To better understand the dynamics of particle flows, we focus on the dynamic behavior of particle flow  (\ref{flow}) as a function of $\lambda$ over $[0, 1]$.

We first introduce a well-known result in the theory of stochastic differential equations - the It\^{o} Lemma.
\vs

{\sc Lemma 4.1 (It\^{o} Lemma).} \cite{Jaz} {\em Let $\phi(x,\lambda)\in \mathbb{R}$ be a scalar-valued real function, continuously differentiable in $\lambda$ and having continuous second mixed partial derivatives with respect to the elements of $x$. Then the stochastic differential $d\phi$ of $\phi$ is
\bq
d\phi = \phi_{\lambda}d\lambda+\nb_x^T\phi dx+\frac{1}{2}tr(Q\nb_x\nb_x^T\phi)d\lambda.
\label{dphi}
\eq
}
\vs
Substituting the form of $dx$ (\ref{flow}) into (\ref{dphi}), we can further write
\bq
d\phi=(L\phi) d\lambda+(\nb_x^T\phi) q(x,\lambda)dw_{\lambda}
\label{dphi_2}
\eq
in which $L$ is the diffusion operator defined as
\[
L\phi = \frac{\partial}{\partial\lambda}\phi+(\nb_x^T\phi)f + \frac{1}{2}tr(Q\nb_x\nb_x^T \phi).
\]
For stability analysis of stochastic differential equations, $L\phi$ plays a role analogous to that of the derivative of a Lyapunov function for the stability analysis of deterministic differential equations \cite{Mao,Sha}. Assuming that the order of taking expectation and differentiation is exchangeable, we know from (\ref{dphi_2}) that the expectation of $\phi$ satisfies \cite{Jaz}
\[ dE[\phi] = E[L\phi]d\lambda.
\] 
In other words, $E[L\phi]$ governs the dynamics of the mean of $\phi$.
\vs

For the stochastic flow (\ref{flow}), it is important to understand changes in the values of the posterior density $p(x,\lambda)$. For example, we would like to avoid the case in which $p(x,\lambda)$ converges to zero, which leads to “particle degeneracy" in particle filtering.
\vs

{\sc Theorem 4.1.} {\em Assume the assumptions (A1) and (A2). Then for the stochastic flow (\ref{flow}), (\ref{f1})-(\ref{q1}), $\log p$ satisfies
\bq
d\log p =  (L\log p)d\lambda+(\nb_x^T\log p)qdw_{\lambda}, \forall \lambda\in[0, 1]
\label{dlogp}
\eq
in which 
\bq
L\log p = \frac{1}{2}(\nb_x^T\log p)Q\nb_x\log p+\ga(\lambda),
\label{Llogp}
\eq
and $\ga(\lambda)$ is independent of $x$ 
\[
\ga(\lambda)=c_h-(b_g+\lambda b_h)^T (\nb_x\nb_x^T\log p)^{-1}b_h
\]
\[
+\frac{1}{2}(b_g+\lambda b_h)^T (\nb_x\nb_x^T\log p)^{-1}(\nb_x\nb_x^T\log h)
(\nb_x\nb_x^T\log p)^{-1}(b_g+\lambda b_h)
+\frac{1}{2}tr(Q\nb_x\nb_x^T\log p)
\]
with $c_h, b_g, b_h$ defined in (\ref{logg}) and (\ref{logh}).
The initial condition is $\log p|_{\lambda=0}=\log g |_{x=x_0}$.
}

{\sc Proof:} Setting $\phi=\log p$, we know from the It\^{o} Lemma that $\log p$ satisfies the following process
\[
d\log p = (L\log p)d\lambda+(\nb_x^T\log p)qdw_{\lambda}
\]
with
\[
L\log p = \log h +\nb_x^T\log p f+\frac{1}{2}tr(Q\nb_x\nb_x^T\log p).
\]
We next show that $L\log p$ has the form (\ref{Llogp}). Substituting the form of $f$ (\ref{f1}) into the previous equation, we obtain
\[
L\log p
=\log h +(\nb_x^T\log p) (\nb_x\nb_x^T\log p)^{-1}[-\nb_x\log h
+K(\nb_x\nb_x^T\log p)^{-1}(\nb_x\log p)]
\]
\[
+\frac{1}{2}tr(Q\nb_x\nb_x^T\log p)
\]
\[
=\log h -(\nb_x^T\log p) (\nb_x\nb_x^T\log p)^{-1}\nb_x\log h 
+\frac{1}{2}(\nb_x^T\log p)Q\nb_x\log p
\]
\[
+\frac{1}{2}(\nb_x^T\log p) (\nb_x\nb_x^T\log p)^{-1}(\nb_x\nb_x^T\log h)
(\nb_x\nb_x^T\log p)^{-1}(\nb_x\log p)
+\frac{1}{2}tr(Q\nb_x\nb_x^T\log p)
\]
\bq
=\frac{1}{2}(\nb_x^T\log p)Q\nb_x\log p+\psi
\label{logp}
\eq
where
\[
\psi=\log h -(\nb_x^T\log p) (\nb_x\nb_x^T\log p)^{-1}\nb_x\log h 
\]
\[
+\frac{1}{2}(\nb_x^T\log p) (\nb_x\nb_x^T\log p)^{-1}(\nb_x\nb_x^T\log h)
(\nb_x\nb_x^T\log p)^{-1}(\nb_x\log p)
+\frac{1}{2}tr(Q\nb_x\nb_x^T\log p).
\]
Note that $\psi$ is quadratic in $x$ under the assumption (A1). Direct verification shows that $\nb_x\psi=0$. Therefore, $\psi$ is a function of $\lambda$ only. We find the value of $\psi$ by setting $x=0$, which leads to
\[
\psi = \psi|_{x=0} = c_h-(b_g+\lambda b_h)^T (\nb_x\nb_x^T\log p)^{-1}b_h
\]
\[
+\frac{1}{2}(b_g+\lambda b_h)^T (\nb_x\nb_x^T\log p)^{-1}(\nb_x\nb_x^T\log h)
(\nb_x\nb_x^T\log p)^{-1}(b_g+\lambda b_h)
+\frac{1}{2}tr(Q\nb_x\nb_x^T\log p)
\]
\[
=\ga(\lambda).
\]
The combination of the previous equation with (\ref{logp}) gives (\ref{Llogp}). Q.E.D.
\vs

In (\ref{Llogp}), the first term of the drift $L\log p$ is non-negative and drives the (mean) value of $\log p$ in increasing direction. The second, $\lambda$-dependent term $\ga(\lambda)$ is a constant that can be readily calculated despite  its complex appearance. Further research is needed to understand $\ga(\lambda)$ to fully evaluate the dynamics of $\log p$. For example, we would like to know analytically, without numerical evaluation for every $\lambda\in [0, 1]$, when $\ga(\lambda)$ is positive or negative.
\vs

We next consider the dynamics of $\nb_x\log p\in\mathbb{R}^n$ as a function of $\lambda$ over $[0, 1]$. For the sake of generality, we tentatively state the results without assuming the assumption (A1).
\vs

{\sc Theorem 4.2.} {\em Assume the assumption (A2). For the stochastic flow (\ref{flow}), (\ref{f1})-(\ref{q1}), $y= \nb_x\log p$ satisfies
\bq
dy = (Ly)d\lambda+(\nb_x\nb_x^T\log p)qdw_{\lambda}, \forall \lambda\in[0, 1]
\label{dlogp_dx}
\eq
with 
\bq
Ly =K(\nb_x\nb_x^T\log p)^{-1}y+\frac{1}{2}\xi
\label{Ly}
\eq
in which $\xi\in\mathbb{R}^n$ with $i$-th component $\xi_i$ being
\[
\xi_i = tr(Q\nb_x\nb_x^T(y_i)).
\]
The initial condition is
\[
y_0\df y|_{\lambda=0}=\nb_x\log g|_{x=x_{0}}. 
\]
}
{\sc Proof:} By setting $\phi$ to be each component of $y$ and applying It\^{o} Lemma, we have
\bq
dy = (Ly) d\lambda +(\nb_x\nb_x^T\log p)qdw_{\lambda}
\label{dy1}
\eq
where
\bq
Ly = \nb_x\log h +(\nb_x\nb_x^T\log p)f+\frac{1}{2}\xi.
\label{Ly_2}
\eq
By substituting (\ref{f1}) into (\ref{Ly_2}) we obtain
(\ref{Ly}). Q.E.D.
\vs

If we further assume the assumption (A1), $y=\nb_x\log p$ is linear in $x$, which leads to $\xi=0$. In this case, the coefficient matrix of (\ref{dlogp_dx}) is
\[
F \df K(\nb_x\nb_x^T\log p)^{-1}=\frac{1}{2}(\nb_x\nb_x^T\log p)Q_1,
\]
where
\[ Q_1=Q+(\nb_x\nb_x^T\log p)^{-1}(\nb_x\nb_x^T\log h)(\nb_x\nb_x^T\log p)^{-1}.
\]
For a unimodal probability distribution (e.g., Gaussian distributions), negative definiteness of the Hessian matrix of its density function is a mild assumption in practice. Assume that  $\nb_x\nb_x^T\log p$ is negative definite. If $Q$ is chosen such that $Q_1$ is positive semi-definite, then $F$ is a product of a negative definite matrix $(1/2)\nb_x\nb_x^T\log p$ and a positive semi-definite matrix $Q_1$. In this case, all the eigenvalues of $F$ are non-positive.  However, we cannot conclude based on eigenvalues only that (\ref{dlogp_dx}) is stable because $F$ is $\lambda$-varying \cite{Kha}. We need to utilize the stability theory for time-varying systems.
We adopt a Lyapunov-like function to characterize the dynamics of particle flows. In particular, we denote $M(\lambda)=(-\nb_x\nb_x^T\log p)^{-1}$,
and define
\bq
V(x,\lambda)=(\nb_x^T\log p) M(\lambda)(\nb_x\log p).
\label{V}
\eq
The reason there is a negative sign in front of $\nb_x\nb_x^T\log p$ in the definition of $M$ is that $\nb_x\nb_x^T\log p$ is typically negative definite, e.g., for non-degenerative Gaussian distributions. With a positive definite matrix $M$, $V(x,\lambda)$ defines a Lyapunov function of $\nb_x\log p$. 
\vs

{\sc Theorem 4.3.} {\em Assume the assumption (A2). Then we have
\bq
dV
=LV d\lambda+(-2\nb_x\log p+u)^T qdw_{\lambda}, \forall \lambda\in [0, 1]
\label{dV}
\eq
in which
\bq
LV = -(\nb_x^T\log p)(Q-U_1)(\nb_x\log p)+tr(Q[M^{-1}+U_2]),
\label{LV}
\eq
with
\[
U_1 = M(\sum_{i=1}^n\frac{\partial \nb_x\nb_x^T\log p}{\partial x_i}f_i)M,
\]
$f_i$ is the $i$-th element of $f$, 
$u\in R^n$ with $i$-th element being
\[
u_i = (\nb_x^T\log p)M(\frac{\partial \nb_x\nb_x^T\log p}{\partial x_i})M(\nb_x\log p),
\]
and 
\[
U_2 =\nb_x^T u.
\]
The initial condition is
\[
V|_{\lambda=0} = (\nb_x^T\log g)(-\nb_x\nb_x^T\log g)^{-1}(\nb_x\log g)|_{x=x(0)}.
\]
}

{\sc Proof:} First, we apply the It\^{o} Lemma to $\phi=V(x,\lambda)$, and also using the notation of diffusion operator $L$,  to obtain
\bq
dV=\nb_{\lambda}V d\lambda+(\nb_x^T V)dx+\frac{1}{2}tr(Q\nb_x\nb_x^TV)d\lambda
=LV d\lambda+(\nb_x^T V)qdw_{\lambda}
\label{DV_2}
\eq
in which
\bq
LV = \nb_{\lambda}V +(\nb_x^T V)f +\frac{1}{2}tr(Q\nb_x\nb_x^TV).
\label{LV_2}
\eq
To prove Theorem 4.3, we need to show that $LV$ has the form (\ref{LV}) and $\nb_x V=-2\nb_x\log p+u$. To that end, we next examine each term in (\ref{LV_2}). For the $V(x,\lambda)$ defined in $(\ref{V})$, using again the equality for differentiating an invertible matrix $B(\theta)$, $dB^{-1}/d\theta=-B^{-1}(dB/d\theta)B^{-1}$,  we have
\bq
\nb_{\lambda}V
=2(\nb_x^T\log h)M(\nb_x\log p)
+(\nb_x^T\log p) M(\nb_x\nb_x^T\log h)M\nb_x\log p
\label{pV1}
\eq
and 
\[
(\nb_x^T V) f = 2(\nb_x^T\log p) M (\nb_x\nb_x^T\log p) f
+\sum_{i=1}^n (\nb_x^T\log p)\frac{\partial M}{\partial x_i}(\nb_x\log p)f_i
\]
\[
= 2(\nb_x^T\log p) M (\nb_x\nb_x^T\log p) f
+\sum_{i=1}^n (\nb_x^T\log p)M\frac{\partial \nb_x\nb_x^T\log p}{\partial x_i}M(\nb_x\log p)f_i
\]
\[
= 2(\nb_x^T\log p) M (\nb_x\nb_x^T\log p) f
+(\nb_x^T\log p)[\sum_{i=1}^n M\frac{\partial \nb_x\nb_x^T\log p}{\partial x_i}M f_i](\nb_x\log p)
\]
\bq
=2(\nb_x^T\log p) M (\nb_x\nb_x^T\log p) f
+(\nb_x^T\log p)U_1(\nb_x\log p)
\label{pV2_1}
\eq
Using the forms of $f$ (\ref{f1}) and $K$  (\ref{q1}), we know from (\ref{pV2_1}) that
\[
(\nb_x^T V) f = 2\nb_x^T\log p M (-\nb_x\log h-KM\nb_x\log p)
\]
\[
=-2\nb_x^T\log h M\nb_x\log p-2\nb_x^T\log p (MKM)\nb_x\log p
\]
\[
=-2\nb_x^T\log h M\nb_x\log p
-2\nb_x^T\log p [\frac{1}{2}Q+\frac{1}{2}M(\nb_x\nb_x^T\log h)M]\nb_x\log p
\]
\bq
=-2\nb_x^T\log h M\nb_x\log p
-\nb_x^T\log p [Q+ M(\nb_x\nb_x^T\log h)M]\nb_x\log p.
\label{pV2}
\eq
Furthermore,
\[
\nb_x V
=2(\nb_x\nb_x^T\log p) M\nb_x\log p+u
=-2\nb_x\log p+u
\]
and consequently
\bq
\nb_x\nb_x^T V=-2\nb_x\nb_x^T\log p+U_2.
\label{pV3}
\eq
Substituting (\ref{pV1}), (\ref{pV2}), and (\ref{pV3}) into (\ref{LV_2}), and consolidating terms, we finally arrive at
\[
LV
=-(\nb_x^T\log p)(Q-U_1)\nb_x\log p
-tr(Q\nb_x\nb_x^T\log p)+tr(QU_2)
\]
which is (\ref{LV}) since $M^{-1}=(-\nb_x\nb_x^T\log p)$. Q.E.D.
\vs

We return to the notation $y=\nb_x\log p$ and define
\bq
V_1(y) =y^T M_1 y, M_1=(-\nb_x\nb_x^T\log g)^{-1}
\label{V1}
\eq
\bq
V_2(y) =y^T M_2 y, M_2=(-\nb_x\nb_x^T\log g -\nb_x\nb_x^T\log h)^{-1}
\label{V2}
\eq
in which $M_1$ and $M_2$ are constant matrices under the assumption (A1) and not functions of $\lambda$. Under the assumption (A3) below, $V_1$ and $V_2$ are Lyapunov functions. We next examine the dynamics of $V(y,\lambda)=V(x,\lambda)$ as $\lambda$ changes over $[0, 1]$. We first consider the Exact Flow that was first derived in \cite{DHN2010}. The Exact Flow corresponds to the case of $Q=0$ in (\ref{q1}).
\vs

{\sc Corollary 4.1.} {\em Consider the Exact Flow with $Q=0$. Assume the assumptions (A1), (A2), and that 
\begin{description}
\item[(A3)] The matrix $\nb_x\nb_x^T\log g$ is negative definite, and $\nb_x\nb_x^T\log h$ is negative semi-definite.
\end{description}
Then $y=\nb_x\log p$ satisfies
\bq
V_1(y)\geq V_1(y_0),
V_2(y) \leq V_1(y_0), \forall \lambda \in [0, 1]
\label{bound_exactflow}
\eq
in which $y_0=\nb_x\log g|_{x=x(0)}$.
}

{\sc Proof:} For the Exact Flow, $Q=0$ which is equivalent to $q(x,\lambda)=0$. Under the assumption (A1), $\nb_x\nb_x^T\log p$ is not a function of $x$. Therefore, $U_1=0$. Then
we know from (\ref{dV}) and (\ref{LV}) that $LV\equiv0$ and
\[
\frac{dV}{d\lambda}\equiv 0, \forall \lambda\in [0, 1].
\]
Therefore, $V(y,\lambda)$ is a constant over $\lambda\in [0, 1]$. Note that, for given $y$, $V_2(y)\leq V(y,\lambda)\leq V_1(y)$ for all $\lambda\in [0, 1]$. Consequently,
\[
V_1(y) \geq V(y,\lambda)=V(y,\lambda)|_{\lambda=0}=V_1(y_0), \forall \lambda \in [0, 1],
\]
and
\[
V_2(y) \leq V(y,\lambda)= V(y,\lambda)|_{\lambda=0}=V_1(y_0),\forall \lambda\in[0, 1],
\]
with initial condition
\[
 \nb_x\log p|_{\lambda=0}=\nb_x\log g |_{x=x(0)}=y_0,
\]
which is exactly (\ref{bound_exactflow}). Q.E.D.
\vs

\begin{figure}[h]
	\centering
	\includegraphics[width=4in]{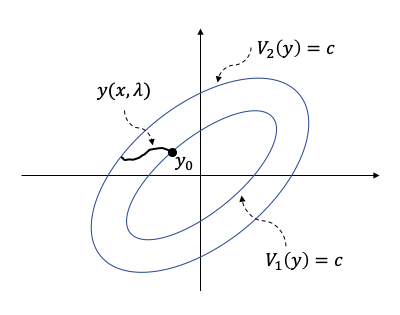}
	\caption{Illustration of the dynamics of $y(x,\lambda)=\nb_x\log p$ for the Exact Flow with $c=V_1(y_0)$}
	\label{fig:fig1}
\end{figure}

Corollary 4.1 states that the dynamics of $y=\nb_x\log p$ is sandwiched between two ellipsoids as $\lambda$ moves from $0$ to $1$, as illustrated by Figure \ref{fig:fig1}. For the Exact Flow, $y$ starts on $V_1(y)=c$ and ends on $V_2(y)=c$. 

\vs

The following Corollary 4.2 summarizes the dynamic behavior for the general case of (\ref{flow}).
\vs

{\sc Corollary 4.2.} {\em Assume the assumptions (A1)-(A3) and that $Q$ is constant (independent of $\lambda$). For the flow $x$ in (\ref{flow}) with positive semi-definite $Q$, 
we can divide the space $\mathbb{R}^n$ into three partitions as the following
\bi
\item 
$S_1\df \{y | y\in\mathbb{R}^n, y^TQy>tr(QM_2^{-1})\}$
\item 
$
S_2\df \{y | y\in\mathbb{R}^n, y^TQy<tr(QM_1^{-1})\}
$
\item 
$
S_3\df \{y | y\in\mathbb{R}^n, tr(QM_1^{-1})\leq y^TQy\leq tr(QM_2^{-1})\}
$
\ei
For $y\in S_1$, $LV$ is decreasing. For $y\in S_2$, $LV$ is increasing. For $y\in S_3$, sign of $LV$ is $\lambda$ dependent.
}

{\sc Proof:} For the general flow $x$ in (\ref{flow}) with positive semi-definite $Q$, the dynamics of $V(y,\lambda)$ is governed by $LV$ in (\ref{LV}).
Under the assumption (A1), $\nb_x\nb_x^T\log p$ is independent of $x$. Consequently, $U_1=0, u=0$, and $U_2=0$. In this case, we have
\bq
LV = -(\nb_x^T\log p)Q(\nb_x\log p)+tr(QM^{-1}).
\label{LV_3}
\eq
Under the assumption (A3), $tr(QM^{-1})\geq 0$ for all $\lambda \in [0, 1]$. Note that $M_1^{-1}\leq M^{-1}\leq M_2^{-1}$. Therefore, we can divide the space $\mathbb{R}^n$ into the three partitions as described. Q.E.D.
\vs

If $Q$ is a matrix function of $\lambda$, and also assume the assumption (A1), (\ref{LV_3}) remains valid. The first term of (\ref{LV_3}) is always non-positive
and the second term non-negative. However, the partitions would be dependent of $Q(\lambda)$. If $Q$ is singular, $S_1, S_2$, and $S_3$ are subspaces.

The dynamics of $y$ is schematically illustrated in Figure 2.
 
 \begin{figure}[h]
 	\centering
 	\includegraphics[width=4in]{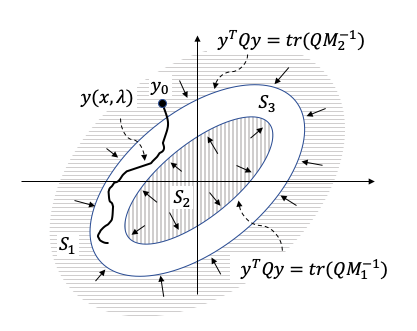}
 	\caption{Illustration of the dynamics of $y(x,\lambda)=\nb_x\log p$ for positive semi-definite $Q$. Short solid arrows indicate moving direction of $V(y,\lambda)$.}
 	\label{fig:fig2}
 \end{figure}

Corollaries 4.1 and 4.2 reveal interesting dynamics of $y=\nb_x\log p$: Assume $y$ is not in the null space of $Q$. If $y$ is too large, the first term in $LV$ dominates, and we thus have $LV<0$ which drives $E[V]$ in decreasing direction. Note that $M(\lambda)$ is bounded, $M_2\leq M(\lambda)\leq M_1$. Therefore, $y$ cannot be too large. On the other hand, if $y$ is too small, the second, positive term in $LV$ dominates. In this case $LV>0$, which drives $E[V]$ in increasing direction, and consequently prevents $y$ from converging to $0$.
\vs

We are interested in $y=\nb_x\log p$ because it is closely related to the maximum likelihood estimate solution.  If $p(x,\lambda)$ is concave and continuously differentiable, solving 
\[
\nb_x\log p = 0
\]
would lead to the maximum likelihood estimate. Corollaries 4.1-4.2 ensure that particles stay close to but do not converge to the maximum likelihood estimate for all $\lambda \in [0, 1]$. Assumption (A3) is mild and satisfied by a wide range of distributions including the Gaussian family.
\vs

The terms $\xi$ in Theorem 4.2 and $u, U_1, U_2$ in Theorem 4.3 capture the non-zero order terms of $x$ in $\nb_x\nb_x^T\log p$. For a particular problem, we may derive their exact forms and subsequently analyze the dynamics of $\nb_x\log p$ and $V(x,\lambda)$. Analysis of (\ref{dV}) without additional assumptions on $\nb_x \nb_x^T\log p$ would be difficult. Further research is needed to identify a particular interesting class of problems, such as the exponential family of distributions, for which we can analyze the dynamical behavior of $V(x,\lambda)$.

\section{Conclusions}
In this paper, we examined dynamic properties of a recently derived family of particle flow filters. We proved that the particle flows indeed have desired posterior distribution, without the Gaussian assumption on the measurement likelihood. We also showed that the particle flows are stable in the sense that the particles stay close to the maximum likelihood estimate, but do not converge to it (to avoid “particle degeneracy"), which is a desirable property for the stability of particle flows. The results guarantee that the particles do not diverge to infinity. As for future efforts, a comprehensive summary of open problems is available in \cite{DH2016}.


\end{document}